\documentclass[sigconf]{acmart}

\usepackage{booktabs} 

\usepackage[ruled]{algorithm2e} 

\SetAlFnt{\small}
\SetAlCapFnt{\small}
\SetAlCapNameFnt{\small}
\SetAlCapHSkip{0pt}

\usepackage{color,xcolor}

\usepackage{colortbl}
\usepackage{float,wrapfig}
\usepackage{multirow}
\usepackage{subcaption}

\usepackage{amsfonts,amsthm}
\usepackage{bm}
\usepackage{bbold}
\usepackage{microtype}

\usepackage{soul}
\usepackage{xspace}

\usepackage{url}

\usepackage{algorithmic}
\usepackage{mathtools}

\newcommand{\ssect}[1]{\S~\ref{#1}}

\newcommand{\fig}[1]{Figure~\ref{#1}}

\newcommand{\tbl}[1]{Table~\ref{#1}}

\newcommand{\ignore}[1]{}

\makeatletter
\DeclareRobustCommand\onedot{\futurelet\@let@token\@onedot}
\def\@onedot{\ifx\@let@token.\else.\null\fi\xspace}

\def\eg{\emph{e.g}\onedot} 
\def\ie{\emph{i.e}\onedot} 
 
\def\etc{\emph{etc}\onedot}

\makeatother

\copyrightyear{2023} 
\acmYear{2023} 
\setcopyright{rightsretained} 
\acmConference[SCF '23]{Symposium on Computational Fabrication}{October 8--10, 2023}{New York City, NY, USA}
\acmBooktitle{Symposium on Computational Fabrication (SCF '23), October 8--10, 2023, New York City, NY, USA}
\acmDOI{10.1145/3623263.3623364}
\acmISBN{979-8-4007-0319-5/23/10}

\begin{document}
\title{Gradient-Based Dovetail Joint Shape Optimization for Stiffness}

\author{Xingyuan Sun}
\affiliation{
 \institution{Princeton University}
 \city{Princeton}
 \state{NJ}
 \country{USA}
}
\email{xs5@princeton.edu}
\author{Chenyue Cai}
\affiliation{
 \institution{Princeton University}
 \city{Princeton}
 \state{NJ}
 \country{USA}
}
\email{cc4880@princeton.edu}
\author{Ryan P. Adams}
\affiliation{
 \institution{Princeton University}
 \city{Princeton}
 \state{NJ}
 \country{USA}
}
\email{rpa@princeton.edu}
\author{Szymon Rusinkiewicz}
\affiliation{
 \institution{Princeton University}
 \city{Princeton}
 \state{NJ}
 \country{USA}
}
\email{smr@princeton.edu}

\begin{abstract}

It is common to manufacture an object by decomposing it into parts that can be assembled.
This decomposition is often required by size limits of the machine, the complex structure of the shape, \etc.\enspace
To make it possible to easily assemble the final object, it is often desirable to design geometry that enables robust connections between the subcomponents.
In this project, we study the task of dovetail-joint shape optimization for stiffness using gradient-based optimization.
This optimization requires a differentiable simulator that is capable of modeling the contact between the two parts of a joint, making it possible to reason about the gradient of the stiffness with respect to shape parameters.
Our simulation approach uses a penalty method that alternates between optimizing each side of the joint, using the adjoint method to compute gradients.
We test our method by optimizing the joint shapes in three different joint shape spaces, and evaluate optimized joint shapes in both simulation and real-world tests.
The experiments show that optimized joint shapes achieve higher stiffness, both synthetically and in real-world tests.

\end{abstract}



\maketitle

\section{Introduction}

\begin{figure}[t]
    \captionsetup[subfloat]{margin=3pt,format=hang,singlelinecheck=false}
    \centering
    \includegraphics[width=0.65\linewidth]{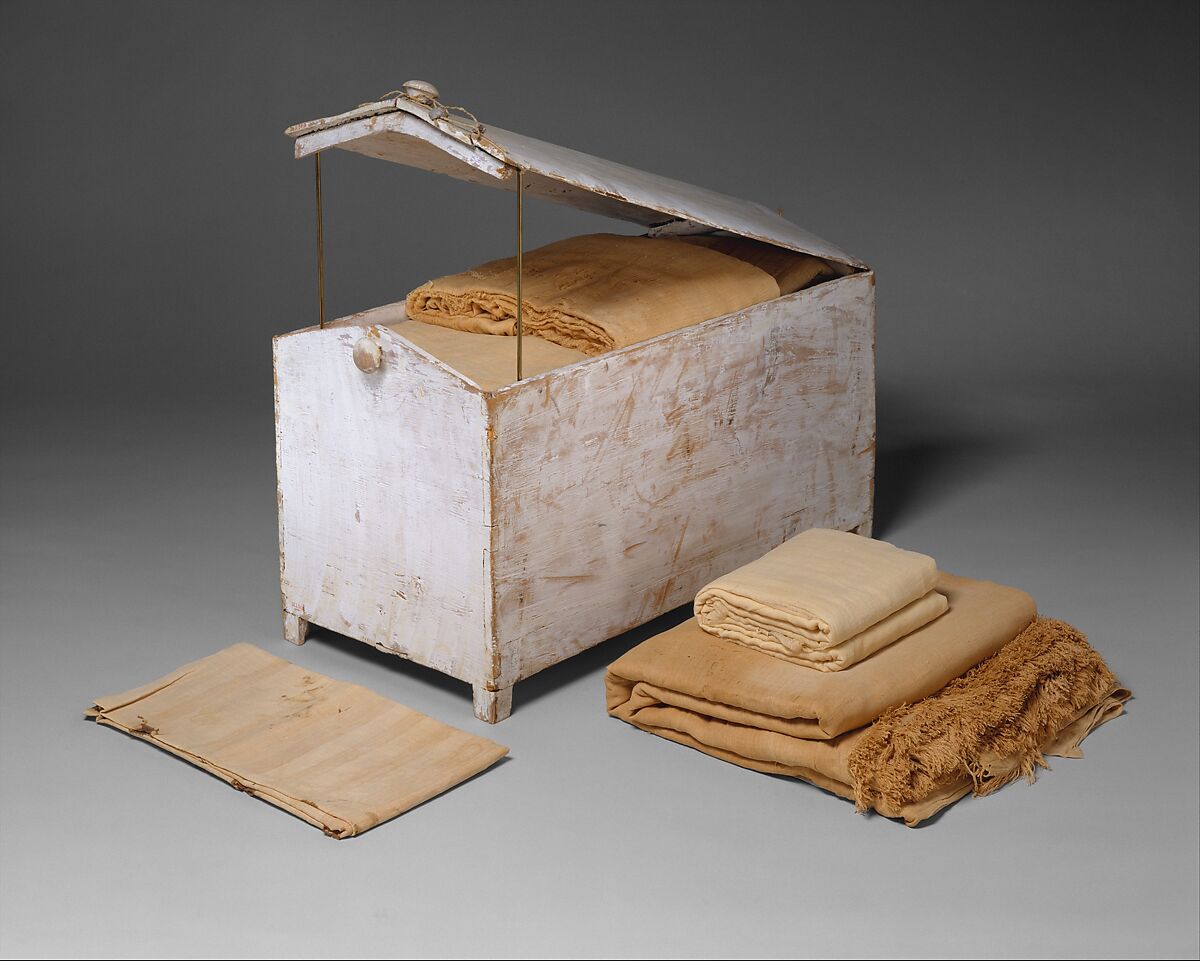}
    \caption{Dovetailed box from Hatnefer's tomb, ca.\ 1492–1473 B.C. (Public domain image via The Metropolitan Museum of Art.)}
    \label{fig:dovetail-box}
\end{figure}

\begin{figure}[t]
    \centering
    \begin{subfigure}{0.49\linewidth}
        \includegraphics[width=\linewidth]{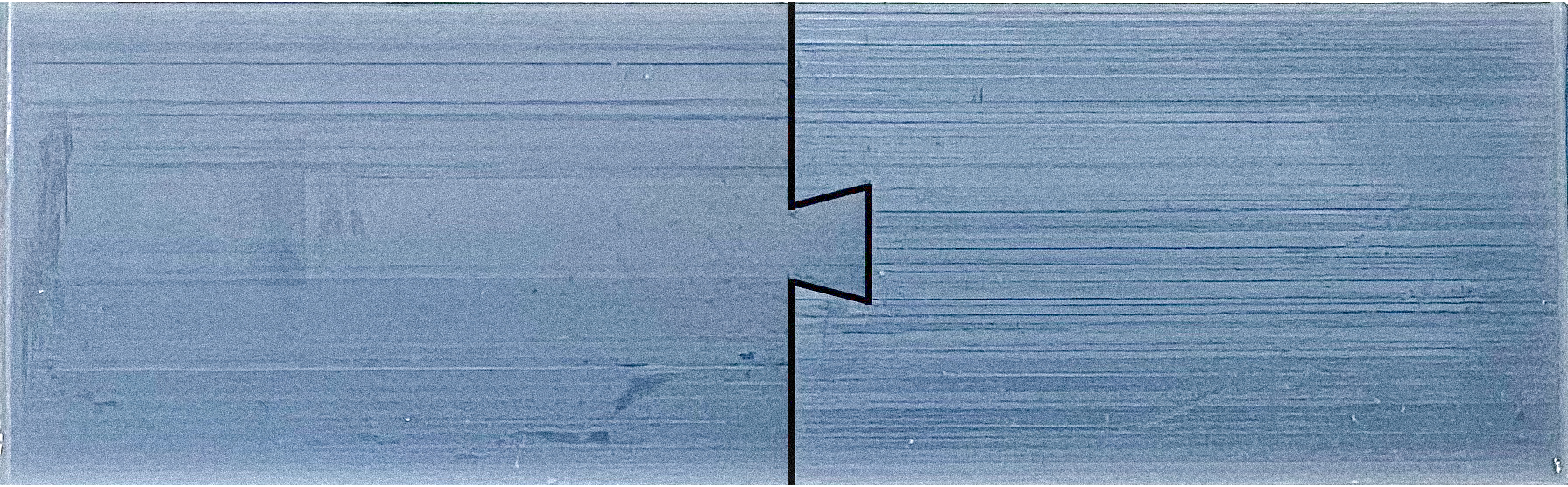}
        \caption{Initial design (single dovetail joint), 228.4 N/mm}
    \end{subfigure}
    \begin{subfigure}{0.49\linewidth}
        \includegraphics[width=\linewidth]{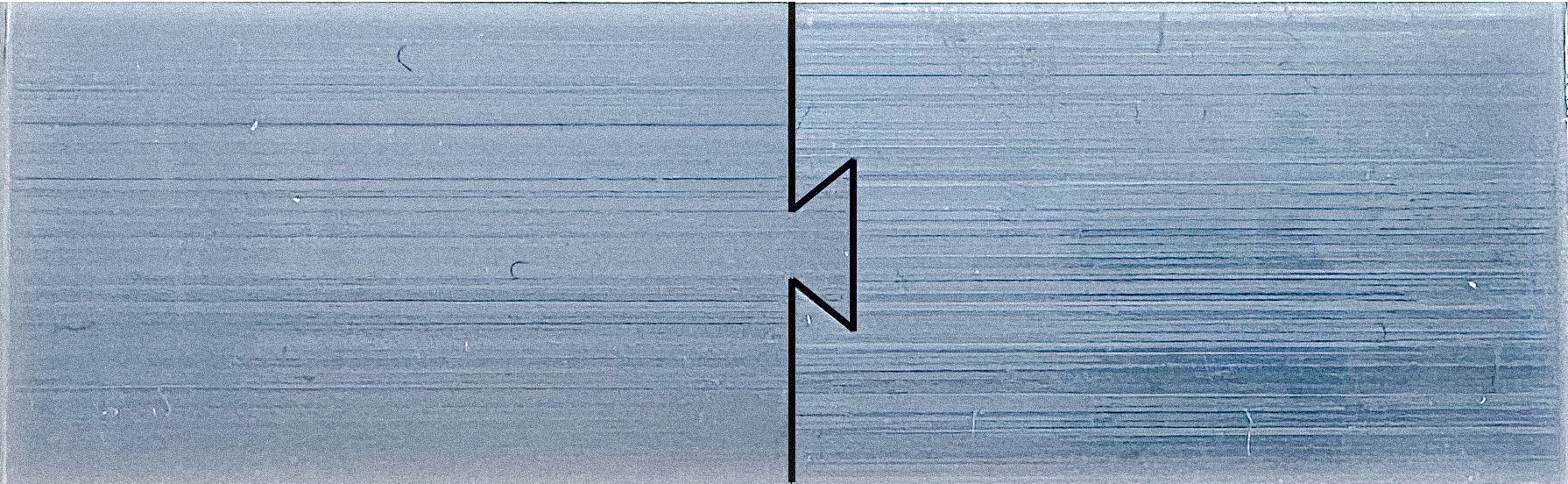}
        \caption{Optimized design (single dovetail joint), 406.0 N/mm}
    \end{subfigure}
    \begin{subfigure}{0.49\linewidth}
        \includegraphics[width=\linewidth]{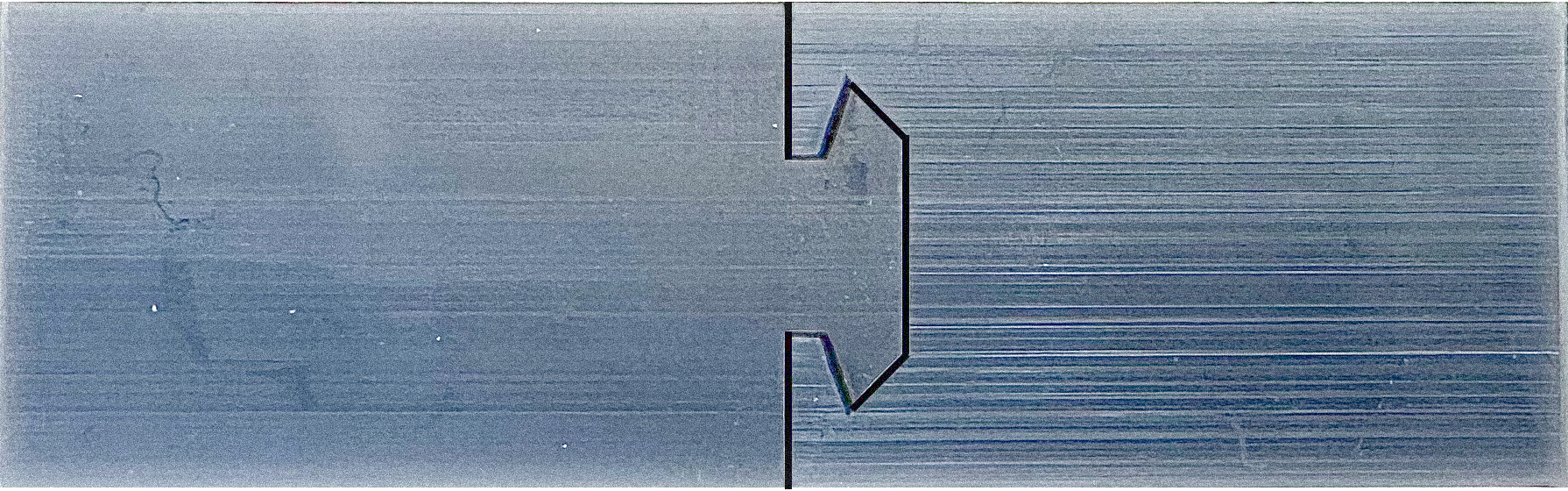}
        \caption{Initial design (complex dovetail joint), 518.1 N/mm}
    \end{subfigure}
    \begin{subfigure}{0.49\linewidth}
        \includegraphics[width=\linewidth]{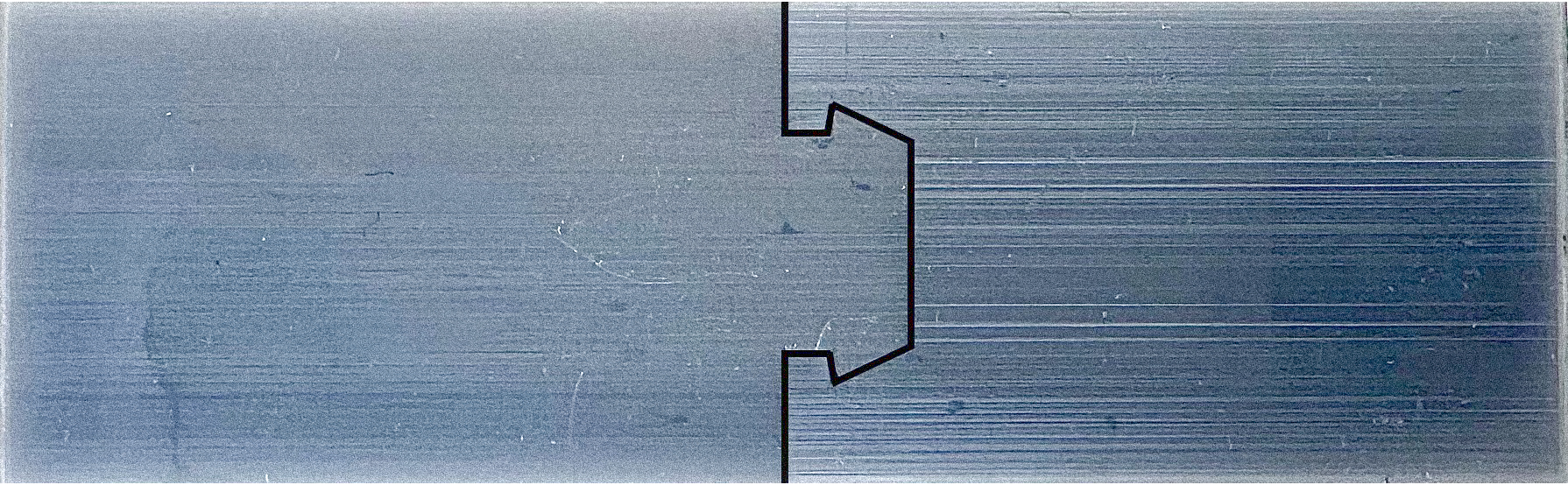}
        \caption{Optimized design (complex dovetail joint), 1036.3 N/mm}
    \end{subfigure}
    \caption{Initial and optimized designs of single and complex dovetail joints with average stiffness measured over three batches and external forces applied on the two sides. (a) (c): initial (randomly chosen) single and complex dovetail joints. (b) (d): optimized single and complex dovetail joints, which provide greater stiffness.}
    \label{fig:p3-teaser}
\end{figure}

The dovetail joint has been used in manufacturing for millennia; Figure~\ref{fig:dovetail-box}, for example, is a dovetailed box from an Egyptian tomb estimated to have been made circa 1492–1473~B.C\@.
In modern manufacturing, dovetail joints~\citep{ruiz1984investigation}, as shown in~\fig{fig:p3-teaser},
are commonly used to connect components in woodworking, turbine blades, 3D printing, \etc.\enspace
In woodworking, there are widely-used heuristics regarding the appropriate angles for hardwood versus softwood \cite{moskowitz2009joiner}, but how can we determine the optimal dovetail shape under more general conditions and with different materials?

In this work, we study the problem of efficiently finding the optimal dovetail joint design that maximizes the stiffness, given a design space and specified external loads.
We choose to study the dovetail joint for its simplicity and ubiquity, but the method we develop should work for any joint shape that is piecewise linear.
To find the optimal design, we formalize the task as an optimization problem, with the stiffness as a function of design parameters that characterize the shape of the dovetail joint in the design space.  We calculate the gradient of stiffness with respect to the design parameters, and directly optimize them using gradient descent.
However, optimizing the shape of a dovetail joint is challenging. First, we need a simulator that can simulate the deformation of the joint given external loads, while considering the contact between two parts of the joint. Second, we need to compute the gradient of stiffness with respect to the design parameters.

To address the first challenge, we perform contact simulations using the penalty approach~\citep{hunvek1993penalty}, alternating between the two sides of the joint.
In every iteration, we compute the deformation of one side of the joint while considering the other side as rigid, and we penalize the elastic side's penetrating the rigid side.
By iteratively applying this alternating approach on the two sides, we have a sufficiently accurate simulator for joint deformation simulation.

As the simulation can be written as partial differential equations (PDEs), the stiffness maximization task can now be understood as a PDE-constrained optimization~\citep{de2015numerical}.
To perform this optimization, we use the adjoint method~\citep{errico1997adjoint, cao2003adjoint} to calculate the gradients of stiffness with respect to design parameters.

To demonstrate the effectiveness of our method, we test it both in simulation and in real tests.
We first verify the accuracy of the gradients by comparing the adjoint method with the finite difference method.
We then run optimization on three different dovetail joint design spaces, each using two different initial designs.
Experiments show that optimized designs provide greater stiffness compared to the initial ones, both in simulation and in real tests on 3D-printed structures.
We finally study the sensitivity of the optimization result with respect to material parameters by optimizing using different Poisson's ratios.
Experiments show that the optimization result is not sensitive to different Poisson's ratios.

\section{Related work}

\subsection{Dovetail joint shape optimization}

Researchers have explored the optimization of dovetail joints using different approaches.
The most straightforward approach is simply testing different design parameters.
\citet{kogo2002application} and \citet{kogo2019application} conducted tensile and shear tests on carbon-carbon composite dovetail joints with different dovetail angles.
\citet{miyauchi2006analysis} tested wooden dovetail joints with different inclinations and base widths.
\citet{jeong2012predicting} tested different wooden dovetail joints for maximum
tension load.
\citet{estenlund2022dovetail} studied the dovetail design for mounting coils on 
rotors by building a simulator and enumerating different dovetail angles.
Another approach for dovetail optimization is applying a gradient-free optimizer on a simulator.
\citet{hu2022study} tested dovetails with different combinations of tenon length, width, thickness, and angle, and studied the effect of each design parameter using the linear model.
\citet{yang20182d} used commercial FEM software for simulation and optimized dovetail shapes for aero-engines using several different gradient-free optimizers.
Some researchers build (differentiable) surrogate models for the simulators and optimize the surrogate models.
For example, \citet{hahn2012design} first optimized design parameters using a surrogate model and further used other gradient-free optimizers to optimize design parameters that are sensitive.
In this work, instead of enumerating, using gradient-free optimizers or surrogate models, we directly optimize dovetail design parameters in the FEM simulation, using the adjoint method to compute the gradients.

\subsection{PDE-constrained optimization}
PDE-constrained optimization is a type of constrained optimization problems whose constraints can be written as partial differential equations reflecting the physics that determine the behavior of a system.
See, \eg,~\citet{de2015numerical} for a comprehensive introduction.
In a PDE-constrained optimization problem, the objective function depends on both the design variable and the state variable, and the constraints between them can be written as PDEs.
There are two major types of approaches to solving PDE constrained optimization problems~\citep{herzog2010algorithms}: \textit{black-box} and \textit{all-at-once}.
\textit{all-at-once} methods treat the design variable and the state variable as being independent during optimization, and researchers may use algorithms such as sequential quadratic programming (SQP)~\cite{boggs1995sequential} to solve the problem.
\textit{black-box} methods treat only the design as the independent variable during optimization, and researchers may use the adjoint method to calculate the total derivative of the objective with respect to the design variable.
See~\citet{givoli2021tutorial} for a tutorial on the adjoint method.
In this work, we formalize the dovetail joint shape optimization problem as a PDE-constrained optimization task and use the adjoint method with gradient descent to solve it.

\section{Method}

\begin{figure*}[t]
    \centering
    \includegraphics[width=\linewidth]{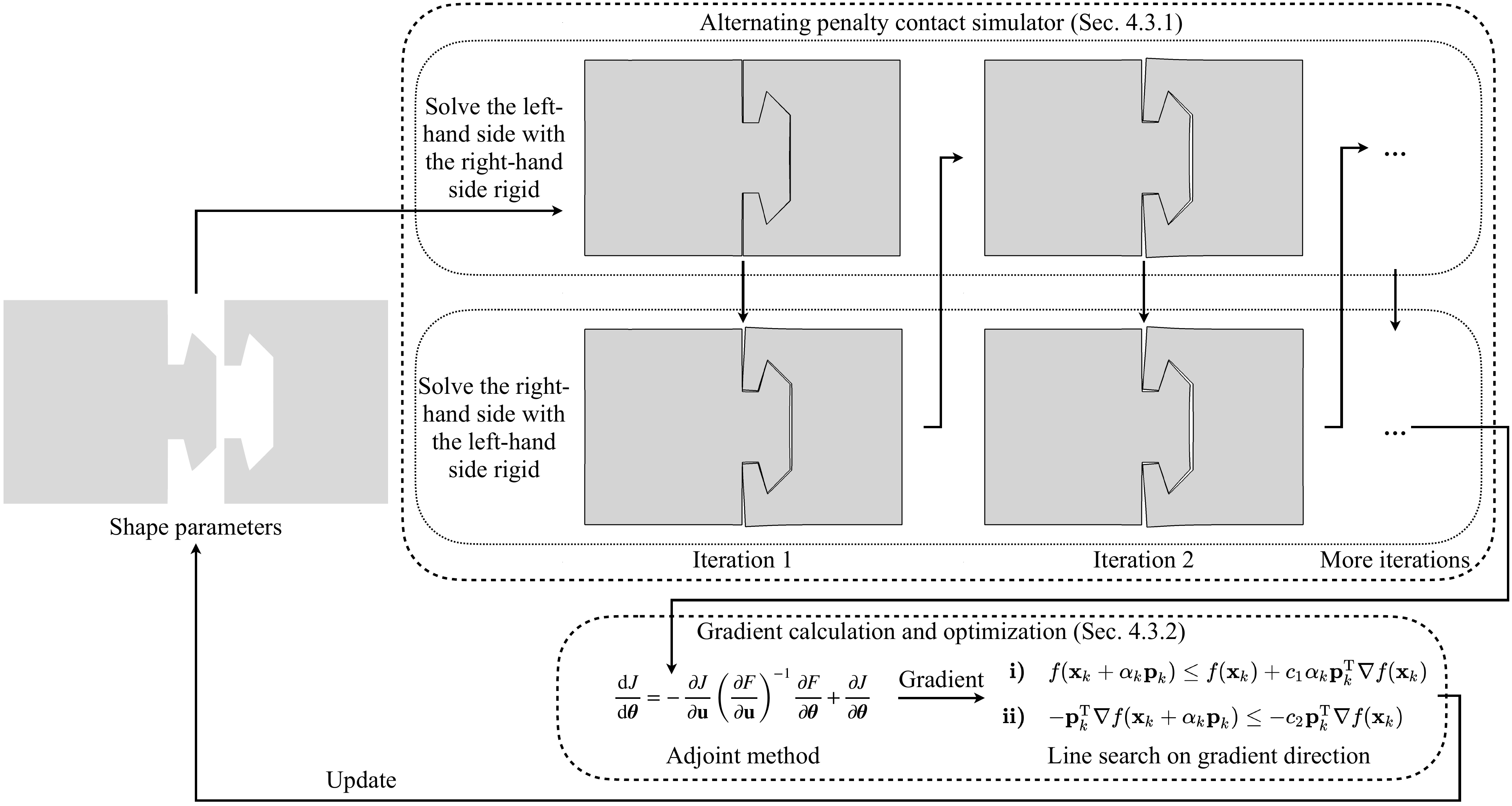}
    \caption{Given a set of shape parameters, we first use the alternating penalty contact simulator (\ssect{ssec:p3-simulator}) to calculate the deformation of the joint, and then use the adjoint method to calculate the gradient and use line search to find the step size for gradient descent (\ssect{ssec:p3-grad-calc-opt}).}
    \label{fig:p3-pipeline}
\end{figure*}

The pipeline of our approach is shown in~\fig{fig:p3-pipeline}.
For an initial set of shape parameters, we first calculate the displacement of the corresponding dovetail joint given a fixed external load on the two sides.
We simulate this by employing a contact solver with the penalty approach, alternating between sides while assuming the other side is rigidly fixed in the deformed configuration of the previous iteration.
After this alternating solver converges, we use the adjoint method to calculate the derivative of the displacement with respect to design parameters and use line search to find an appropriate step size in the direction of steepest descent.

\subsection{Alternating penalty contact simulator}
\label{ssec:p3-simulator}

\begin{figure*}[t]
    \centering
    \begin{subfigure}{0.33\linewidth}
        \includegraphics[width=\linewidth]{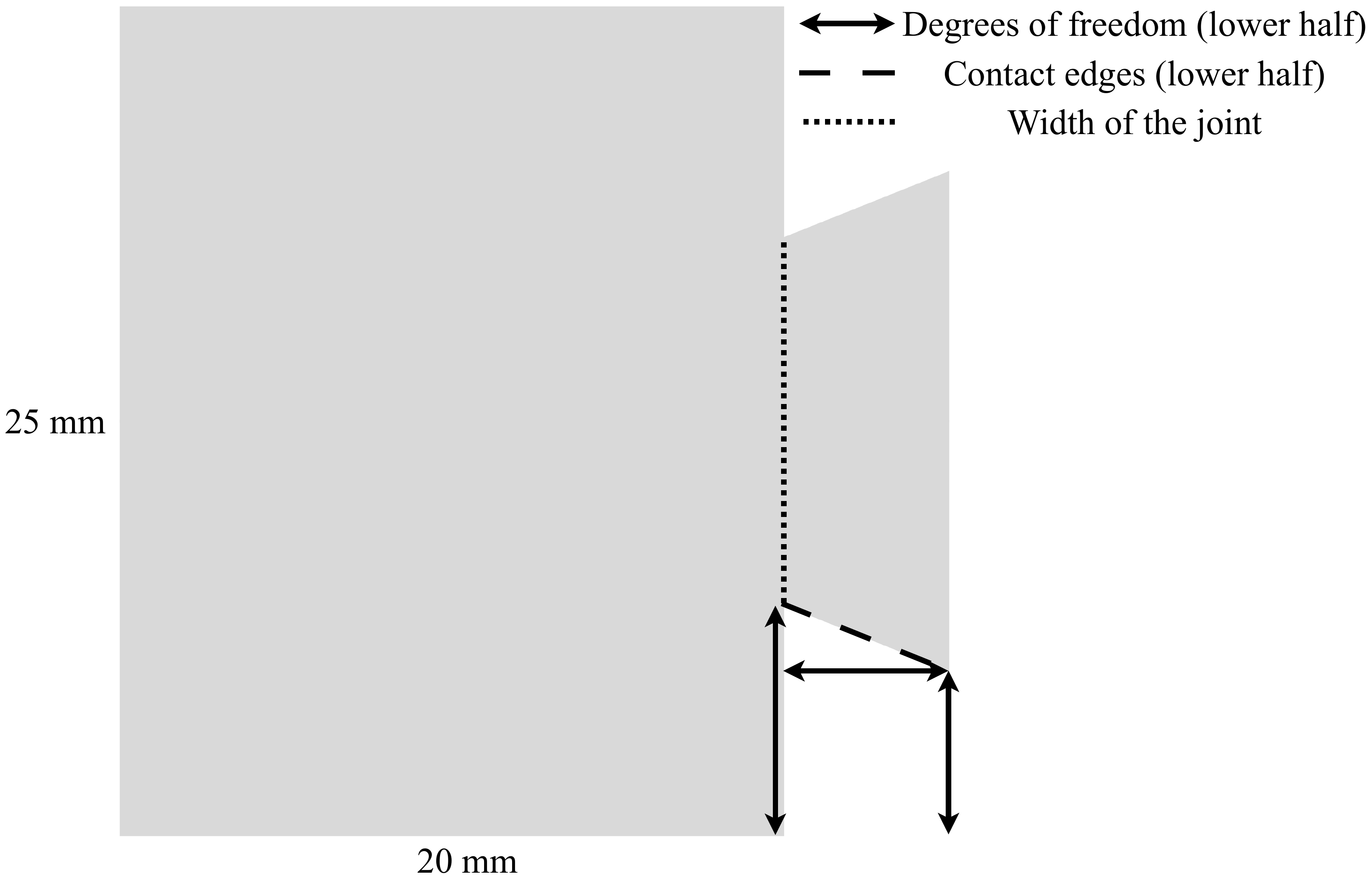}
        \caption{Single dovetail joint}
        \label{fig:p3-single-dovetail}
    \end{subfigure}
    \begin{subfigure}{0.33\linewidth}
        \includegraphics[width=\linewidth]{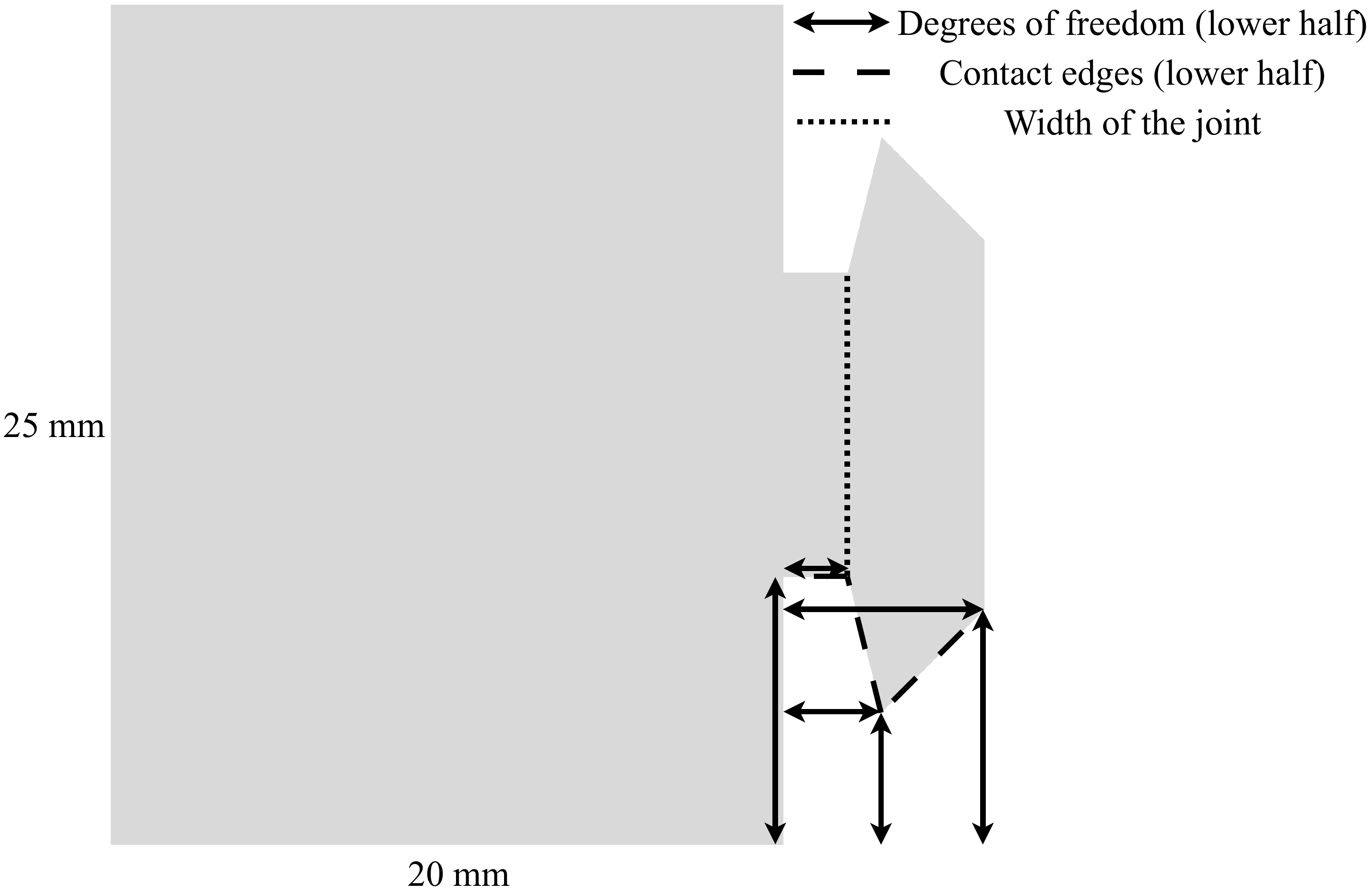}
        \caption{Complex dovetail joint}
        \label{fig:p3-complex-dovetail}
    \end{subfigure}
    \begin{subfigure}{0.33\linewidth}
        \vspace{1em}
        \includegraphics[width=\linewidth]{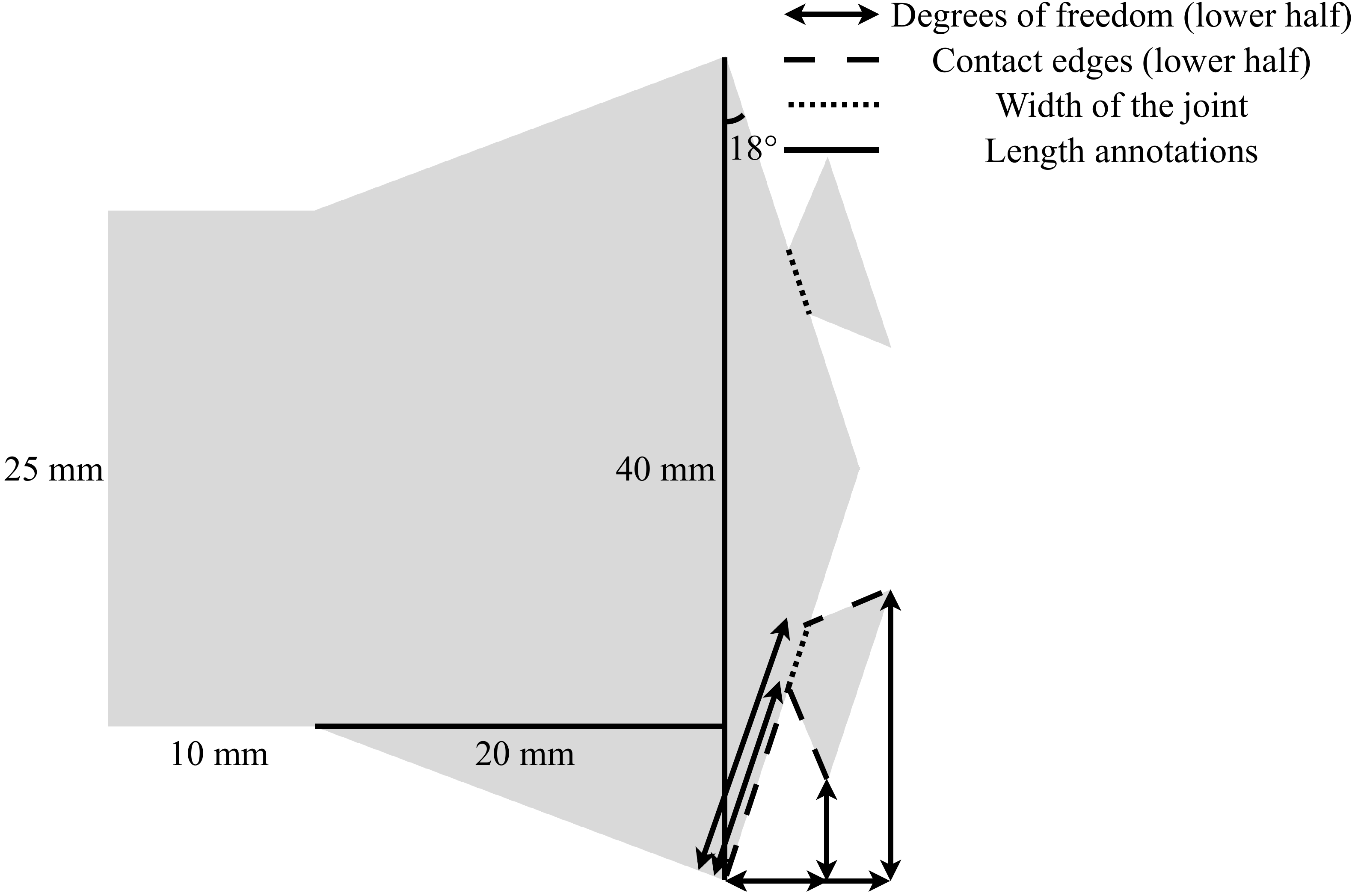}
        \caption{Double dovetail joint}
        \label{fig:p3-double-dovetail}
    \end{subfigure}
    \caption{Three dovetail joint shape design spaces that are used in the experiments. All designs are top-to-bottom symmetric. (a) \textit{single dovetail joint} is the simplest design; three degrees of freedom. (b) \textit{complex dovetail joint} with a design that is more complex; six degrees of freedom. (c) \textit{double dovetail joint} with two dovetails and a non-vertical boundary in the middle; six degrees of freedom. All contact edges and the width are annotated for every design space.}
    \label{fig:p3-shapes}
\end{figure*}

In this subsection, we describe how we simulate the deformation of a specific joint given external loads.
Assuming the material is isotropic, we use the linear elastic model from~\citet{langtangen2017solving}.
Denoting the body as $\Omega$, we have the equations governing the deformation on $\Omega$ as
\begin{align}
\begin{split}
    -\nabla \cdot \bm{\sigma} & = f, \\
    \bm{\varepsilon} & = \frac{1}{2} \bigl(\nabla \mathbf{u} + (\nabla \mathbf{u})^{\!\intercal}\!\bigr), \\
    \bm{\sigma} & = \lambda \text{tr}(\bm{\varepsilon}) I + 2 \mu \bm{\varepsilon},
\end{split}
\label{eqn:p3-linear_elasticity}
\end{align}
where $\bm{\sigma}$ is the stress tensor, $f$ is the body force (0 in our case), $\bm{\varepsilon}$ is the strain tensor, $\mathbf{u}$ is the displacement vector, $\lambda$ and $\mu$ are Lamé parameters, and $I$ is the identity matrix.
Under the assumption of planar stress, we have $\lambda = \frac{E \nu}{1 - \nu^2}$ and $\mu = \frac{E}{2(1 - \nu^2)}$ where $E$ is the Young's modulus of the material (we set it to 1 GPa), and $\nu$ is the Poisson's ratio of the material (we set it to 0.4).
Equivalently, we are minimizing the total potential energy $\Pi$ which is defined as~\cite{alnaes2015fenics}:
\begin{equation}
    \Pi \coloneqq \int_{\Omega} \frac{1}{2} \bm{\varepsilon} : \bm{\sigma} \mathrm{d} A - \int_{\Omega} f \cdot \mathbf{u} \mathrm{d} A - \int_{\partial \Omega} T \cdot \mathbf{u} \mathrm{d} x,
\end{equation}
where the colon is the dot product between tensors and $T$ is the traction force. For reasonable deformations, we set $T$ to 0.001 GPa for single dovetail joints and 0.003 GPa for complex and double dovetail joints (see~\fig{fig:p3-shapes}). We apply equal traction forces on two sides of the joint as Neumann boundary conditions.

One difficulty is to model the contact between the two sides of the joint.
As in~\citet{bleyer2018numericaltours}, we use a penalty approach---solving one side while assuming the other side is rigid, and applying the penalty directly on the displacement field $\mathbf{u}$.
Denote the two sides of the joint as $\Omega_L$ and $\Omega_R$, and consider the case that we want to solve the deformation on $\Omega_L$ while considering $\Omega_R$ rigid.
For simplicity and due to the piecewise-linear boundary of the joints, we fit lines to all the contacting edges (see~\ssect{ssec:p3-shapes} for more details) of $\Omega_R$ and penalize $\mathbf{u}$ on $\Omega_L$ if it collides with the fitted lines.
We have our penalized total potential energy $\Pi_L$ for $\Omega_L$ as
\begin{align}
    \small
    \Pi_L \coloneqq & \int_{\Omega_L} \frac{1}{2} \bm{\varepsilon} : \bm{\sigma} \mathrm{d} A - \int_{\Omega_L} f \cdot \mathbf{u} \mathrm{d} A - \int_{\partial \Omega_L} T \cdot \mathbf{u} \mathrm{d} x \\
    & + w_\texttt{pen} \cdot \int_{\partial \Omega_L} \texttt{softplus}^2(-\texttt{sdf}(\textbf{u}; \Omega_L, \Omega_R)) \mathrm{d} x,
\end{align}
where $w_\texttt{pen}$ is the weight of the penalization term, which we set to 1, $\texttt{softplus}(x) = \left( \ln(1 + \exp(kx)) / k \right)^2$ and $k$ is a scale factor that we set to 50, $\texttt{sdf}$ is the signed distance function and $\texttt{sdf}(\textbf{u}; \Omega_L, \Omega_R)$ measures the signed distance from the deformed left-hand side to the fitted lines of the deformed right-hand side (positive if outside and negative if inside).
We create meshes with a mesh step size of 0.5 mm using pygmsh~\citep{schlomer3pygmsh} and implement the simulator using FEniCS~\citep{alnaes2015fenics}.
We find that four iterations on both sides are usually enough for $\mathbf{u}$ to converge.
Note that the simulator works for any piecewise linear joints, not only for dovetail joints.

\subsection{Gradient calculation and optimization}
\label{ssec:p3-grad-calc-opt}

In this subsection, we describe how we optimize the shape design parameters $\theta$.
With the simulator, we can compute the displacement field $\mathbf{u}(\theta)$.
We define the length change of the joint $d(\mathbf{u}(\theta))$ as the difference of average displacement in the horizontal direction between the left and right edges of the joint.
Note that maximizing the stiffness is equivalent to minimizing the displacement given fixed traction.
We define the optimization objective function $\mathcal{L}(\theta)$ as
\begin{equation}
    \mathcal{L}(\theta) \coloneqq d(\mathbf{u}(\theta)) + w_{\text{min\_l}} \cdot \mathcal{L}_{\text{min\_l}}(\theta) + w_{\text{min\_w}} \cdot \mathcal{L}_{\text{min\_w}}(\theta),
\end{equation}
where $\mathcal{L}_{\text{min\_l}}(\cdot)$ and $\mathcal{L}_{\text{min\_w}}(\cdot)$ are regularizers, and $w_{\text{min\_l}}$ and $w_{\text{min\_w}}$ are weights for the regularizers, both of which we set to 1.
The minimum contact length regularizer penalizes contact between edges that are too short:
\begin{equation}
    \mathcal{L}_{\text{min\_l}}(\theta) \coloneqq \sum_{l \in \texttt{contact}(\theta)} \bigl( \max(\texttt{min\_len} - |l|, 0) \bigr)^2,
\end{equation}
where $\texttt{contact}(\theta)$ is the set of all contacting edges (see~\ssect{ssec:p3-shapes} for more details), $|\cdot|$ measures the length of an edge, and we set $\texttt{min\_len}$ to 1.5 mm.
The minimum width regularization penalty prevents the joint from being too small:
\begin{equation}
    \mathcal{L}_{\text{min\_w}}(\theta) \coloneqq \bigl( \max(\texttt{min\_width} - \texttt{width}(\theta), 0) \bigr)^2,
\end{equation}
where $\texttt{width}(\cdot)$ measures the width of a joint (see~\ssect{ssec:p3-shapes} for more details), and we set $\texttt{min\_width}$ to 3.5 mm.

To calculate the gradient of the objective function with respect to the design parameters, \ie $\frac{\mathrm{d} \mathcal{L}(\theta)}{\mathrm{d} \theta}$, we use the adjoint method, which provides the following result:
\begin{equation}
    \frac{\mathrm{d} \mathcal{L}(\theta)}{\mathrm{d} \theta} = - \frac{\partial \mathcal{L}(\theta)}{\partial \mathbf{u}}\left( \frac{\partial F}{\partial \mathbf{u}} \right)^{-1} \frac{\partial F}{\partial \theta} + \frac{\partial \mathcal{L}(\theta)}{\partial \theta},
\end{equation}
where ${F(\mathbf{u}, \theta) = 0}$ is the PDE corresponding to the simulator.
We implement the automatic gradient calculation using dolfin-adjoint~\citep{mitusch2019dolfin, dokken2020automatic} and PyTorch~\citep{paszke2019pytorch}.

To optimize $\theta$, we use gradient descent.
For every step, we perform a line search along the gradient direction using SciPy~\citep{virtanen2020scipy}, finding a step size that satisfies strong Wolfe conditions~\citep{wolfe1969convergence, wolfe1971convergence}.
If the line search fails, we randomly sample a step size from $\mathcal{N}(0, 0.5^2)$, where a negative step size indicates moving in the opposite direction. This approach introduces controlled noise to prevent entrapment in local minima.
We perform 15 optimization steps and keep the step with minimum $d(\mathbf{u}(\theta))$.
Lastly, to prevent landing in design parameters that are sensitive to manufacturing errors, every time we evaluate the objective function or its gradient, we apply independent random noise sampled from $\mathcal{N}(0, 0.01^2)$ to every dimension of $\theta$ three times and take the average.

\section{Experiments}

In this section, we study the effectiveness of our method. We present different dovetail joint design spaces for all the experiments in~\ssect{ssec:p3-shapes}.
We then provide simulated results (\ssect{ssec:p3-simulator-results}) and check the correctness of gradients on mesh vertex coordinates (\ssect{ssec:p3-grad-check}).
We show the main result---the optimization results in \ssect{ssec:p3-optimization}, evaluating them both synthetically and in real experiments.
Finally, to better understand the effect of material parameters, we show the sensitivity test of optimization results with respect to the Poisson's ratio in \ssect{ssec:p3-poisson}.

\subsection{Shapes for experiments}
\label{ssec:p3-shapes}

As shown in~\fig{fig:p3-shapes}, we use three different joint design spaces for all the experiments.
We only visualize the left part of the joint, as the right part is complementary to it.
All joints are top-to-bottom symmetric, so all simulation is only performed on the lower half of the mesh for lower computational cost.
The first design space is named \textit{single dovetail joint} (\fig{fig:p3-single-dovetail}), which contains the simplest form of a dovetail joint and three degrees of freedom.
The second design space is called \textit{complex dovetail joint} (\fig{fig:p3-complex-dovetail}), which still contains only one dovetail shape but with a structure that is more complex and has six degrees of freedom.
The third design space is \textit{double dovetail joint} (\fig{fig:p3-double-dovetail}), which contains two dovetail structures, a non-vertical boundary in the middle, and six degrees of freedom.
We annotate all the contacting edges for each design space, which are used in the contact penalizer and the minimum contact length regularizer.
We also label the width of every design space, which is used in the minimum width regularize.

\subsection{Alternating penalty contact simulator results}
\label{ssec:p3-simulator-results}
\begin{figure*}[t]
    \centering
    \includegraphics[width=\linewidth]{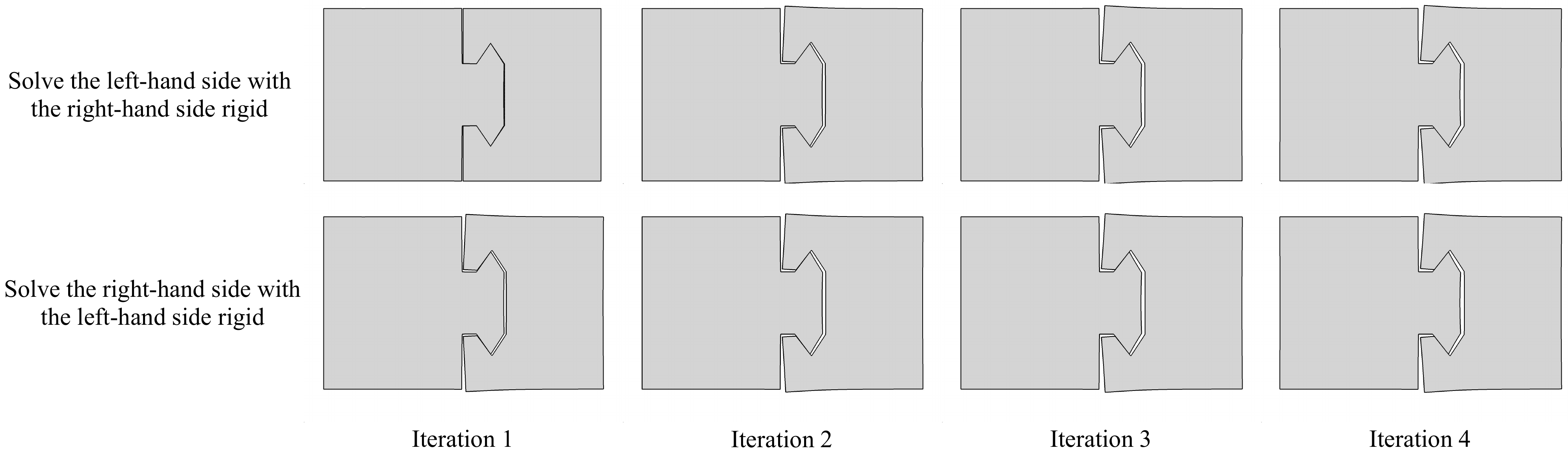}
    \caption{Simulation results on a specific dovetail joint design. The alternating simulator produces reasonable results as the number of iterations increases, and the results converge in the end.}
    \label{fig:p3-simulator-complex-dovetail}
\end{figure*}

In this subsection, we visualize simulation results from the alternating penalty contact simulator to demonstrate it is producing reasonable results, and the results are shown in~\fig{fig:p3-simulator-complex-dovetail}. In the first iteration, we undertake the two-step process: first, solving the left-hand side while keeping the right-hand side fixed, leading to a slight leftward movement of the left side. Subsequently, in the same iteration, we address the right-hand side while maintaining the left side in a fixed state, causing the dovetail section on the right side to bend in opposite directions. In the second iteration, we observe a progressive leftward pull of the left-hand side when the right-hand side remains fixed, followed by increased bending of the right-hand side, consequently creating more space for the left-hand side to experience a further leftward pull in the third iteration. With each iteration, the results evolve and eventually converge as the number of iterations increases.

\subsection{Gradient correctness check}
\label{ssec:p3-grad-check}
\begin{figure*}[t]
    \captionsetup[subfloat]{margin=3pt,format=hang,singlelinecheck=false}
    \centering
    \begin{subfigure}{0.24\linewidth}
        \includegraphics[width=\linewidth]{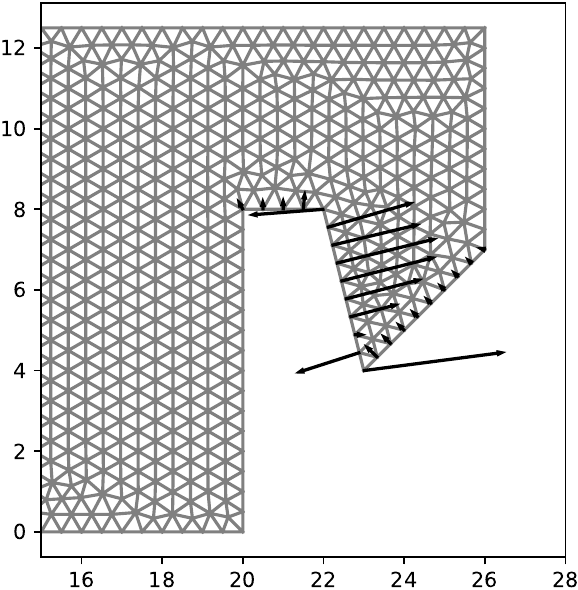}
        \caption{Gradients on the left half; adjoint method}
    \end{subfigure}
    \begin{subfigure}{0.24\linewidth}
        \includegraphics[width=\linewidth]{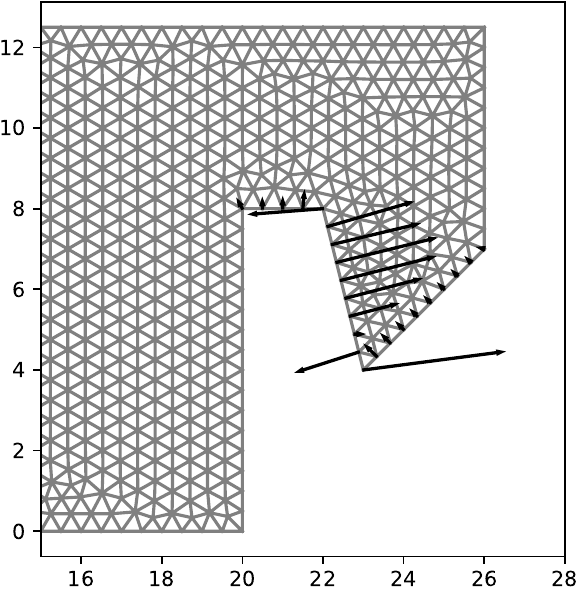}
        \caption{Gradients on the left half; finite difference}
    \end{subfigure}
    \begin{subfigure}{0.24\linewidth}
        \includegraphics[width=\linewidth]{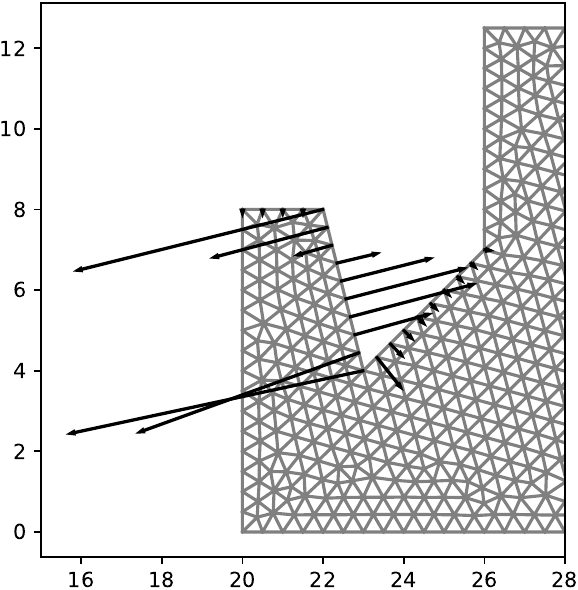}
        \caption{Gradients on the right half; adjoint method}
    \end{subfigure}
    \begin{subfigure}{0.24\linewidth}
        \includegraphics[width=\linewidth]{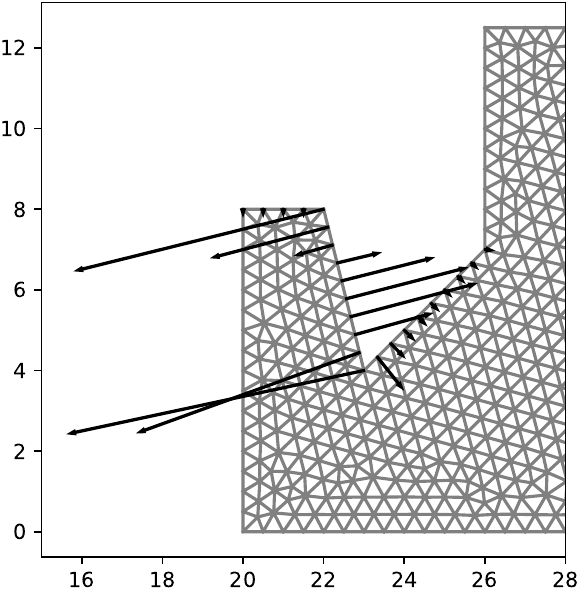}
        \caption{Gradients on the right half; finite difference}
    \end{subfigure}
    \caption{Gradient directions on three contacting edges calculated using the adjoint method and the finite difference method. The two results are indistinguishable, which indicates the gradient calculation is correct.}
    \label{fig:p3-grad-test}
\end{figure*}

In this subsection, we check the correctness of our gradient computation.
We calculate the derivative of the displacement with respect to the coordinates of vertexes from the mesh in two different approaches: the adjoint method and the finite difference method, as shown in~\fig{fig:p3-grad-test}.
For the finite difference method, we use a step size of $10^{-4}$.
We only visualize gradients on three edges, as interior gradients should be all zero.
The two sets of gradients are indistinguishable, which infers our gradient calculation is consistent with the finite difference method.
The average relative difference from the adjoint method results to the finite difference results is $1.82 \times 10^{-4}$, which is negligibly small.

\subsection{Shape optimization for stiffness}
\label{ssec:p3-optimization}

\paragraph{Fabrication and experimental setup.}
\begin{figure*}[t]
    \centering
    \begin{subfigure}{0.33\linewidth}
        \includegraphics[width=\linewidth]{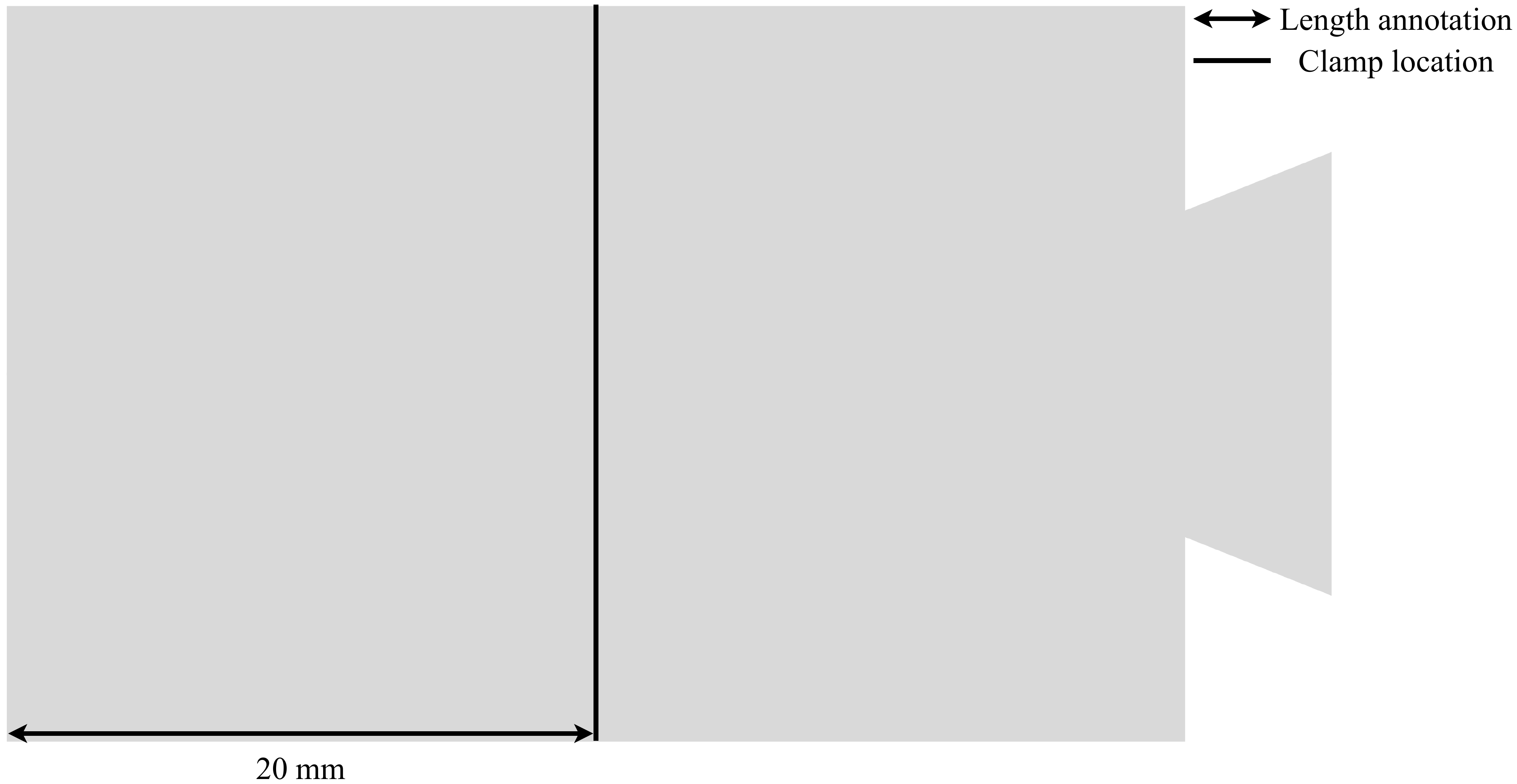}
        \caption{Single dovetail joint}
    \end{subfigure}
    \begin{subfigure}{0.33\linewidth}
        \includegraphics[width=\linewidth]{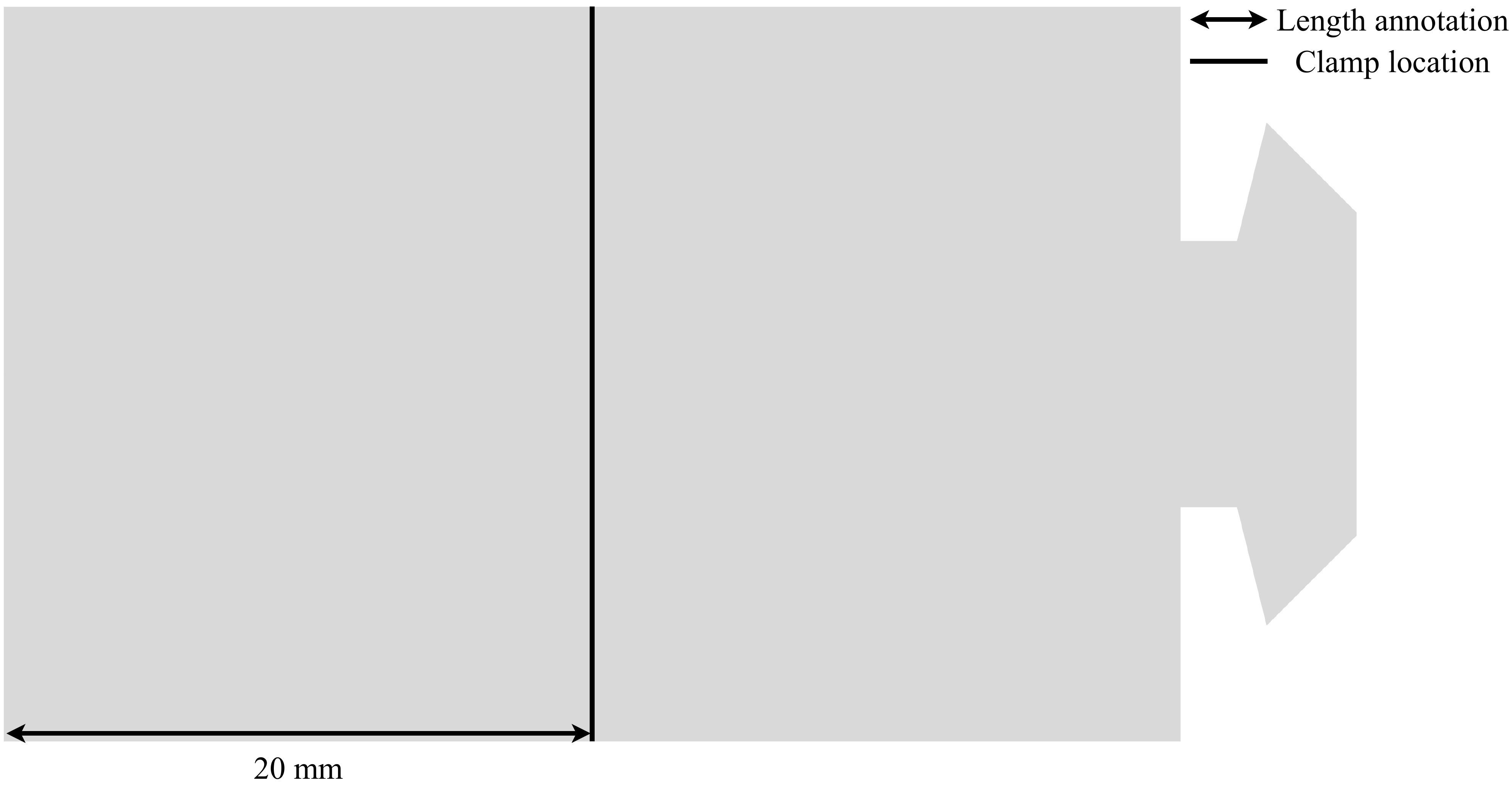}
        \caption{Complex dovetail joint}
    \end{subfigure}
    \begin{subfigure}{0.33\linewidth}
        \includegraphics[width=\linewidth]{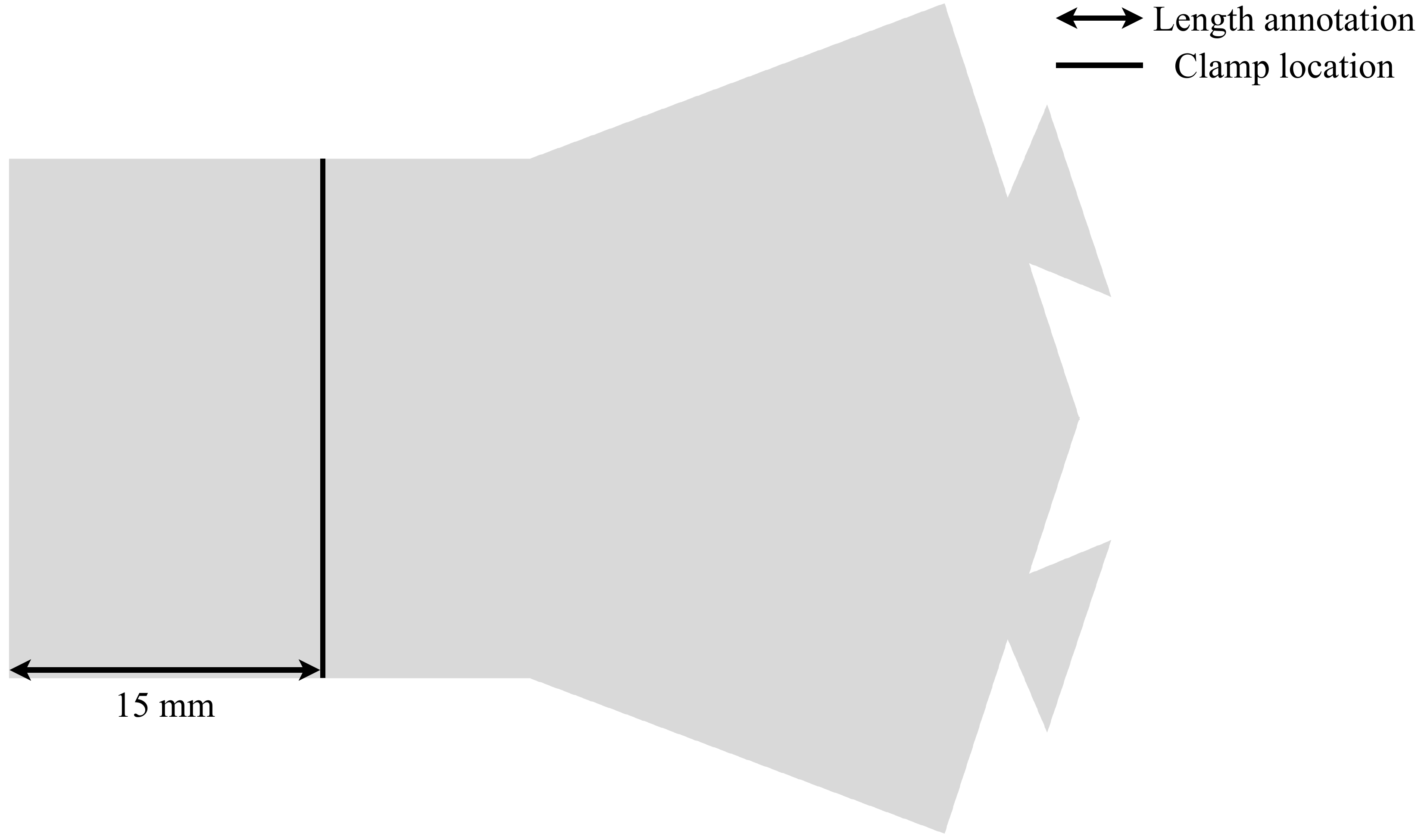}
        \caption{Double dovetail joint}
    \end{subfigure}
    \caption{Tabs are added in real printings such that the universal testing machine can clamp the printed parts.}
    \label{fig:p3-print-shapes}
\end{figure*}

We use Fusion 360\footnote{\url{https://www.autodesk.com/products/fusion-360}} to draw 3D shapes and CHITUBOX\footnote{\url{https://www.chitubox.com/}} and UVtools\footnote{\url{https://github.com/sn4k3/UVtools}} for slicing.
We use the ELEGOO Saturn S resin 3D printer to print laminates of ABS-like resin with a height of 5 mm.
For the printed parts to be assemblable, we introduce a gap of 0.1 mm between the joints.
To measure the stiffness of a joint, we use an Instron 600DX universal testing machine, which is set to move at a speed of 20 mm/min until the tested object breaks and produces a position-load curve.
We also include tabs on two sides of the joint for the machine to clamp, as shown in~\fig{fig:p3-print-shapes}: the parts on the left-hand side of the solid line are the tabs, and the solid lines indicate the position of the clamp.

\paragraph{Optimization and test results.}
\begin{figure*}[t]
    \centering
    \includegraphics[width=\linewidth]{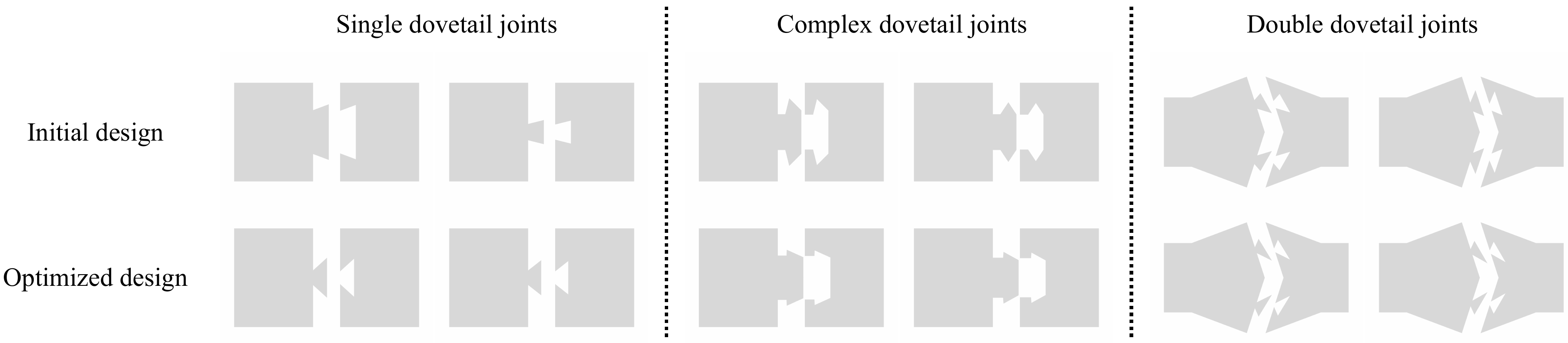}
    \caption{Initial and optimized designs of dovetail joints. The optimized designs are similar for the same design space though the initializations are very different.}
    \label{fig:p3-opt-results}
\end{figure*}

\begin{table}[t]
    \caption{Real measured stiffness of initial and optimized designs over three batches. The significant increase in stiffness indicates that our algorithm successfully optimizes the joint shape design.}
    \centering
    \begin{tabular}{ccc}
        \toprule
        Joint type & Initial design & Optimized design \\
        \midrule
        \multirow{2}{*}{Single dovetail joint} & 219.9$\pm$22.6 & \textbf{362.3$\pm$26.9} \\
        & 228.4$\pm$53.2 & \textbf{406.0$\pm$15.1} \\
        \midrule
        \multirow{2}{*}{Complex dovetail joint} & 518.1$\pm$29.2 & \textbf{1036.3$\pm$61.6} \\
        & 553.3$\pm$55.7 & \textbf{1120.0$\pm$102.4} \\
        \midrule
        \multirow{2}{*}{Double dovetail joint} & 380.6$\pm$23.8 & \textbf{527.4$\pm$48.3} \\
        & 360.2$\pm$12.1 & \textbf{611.8$\pm$64.0} \\
        \bottomrule
    \end{tabular}
    \label{tbl:p3-real-tests}
\end{table}

\begin{figure*}[t]
    \centering
    \includegraphics[width=0.72\linewidth]{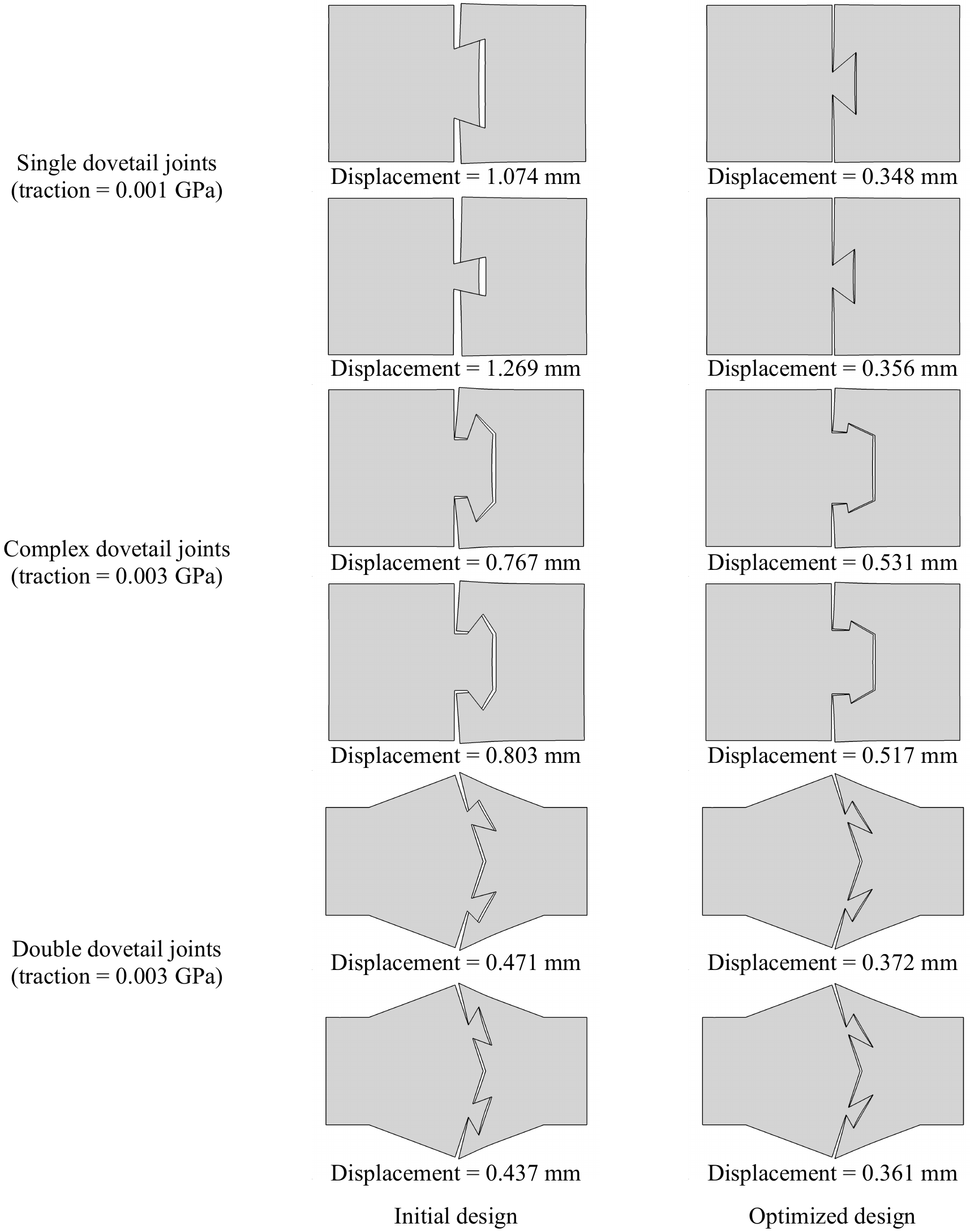}
    \caption{Simulated results and displacements of initial and optimized designs. The displacements of the optimized designs are much smaller than those of the initial designs, which indicates the optimization algorithm is working effectively.}
    \label{fig:p3-synthetic-tests}
\end{figure*}

For each design space, we randomly select two different initial sets of shape parameters and optimize them for fifteen gradient descent steps.
We select the iteration with the smallest simulated displacement, and the results are shown in~\fig{fig:p3-opt-results}.
The optimized results are similar though the initializations are quite different.
The simulated results are shown in~\fig{fig:p3-synthetic-tests}.
We print three copies of each design and test them on the universal testing machine with position-load curves recorded.
We measure the stiffness using the position change between a load of 30 N and a load of 60 N, and the results are listed in~\tbl{tbl:p3-real-tests}.
The significant differences between the initial and optimized designs indicate the effectiveness of our algorithm.

\subsection{Poisson's ratio sensitivity test}
\label{ssec:p3-poisson}

\begin{figure*}[t]
    \centering
    \includegraphics[width=\linewidth]{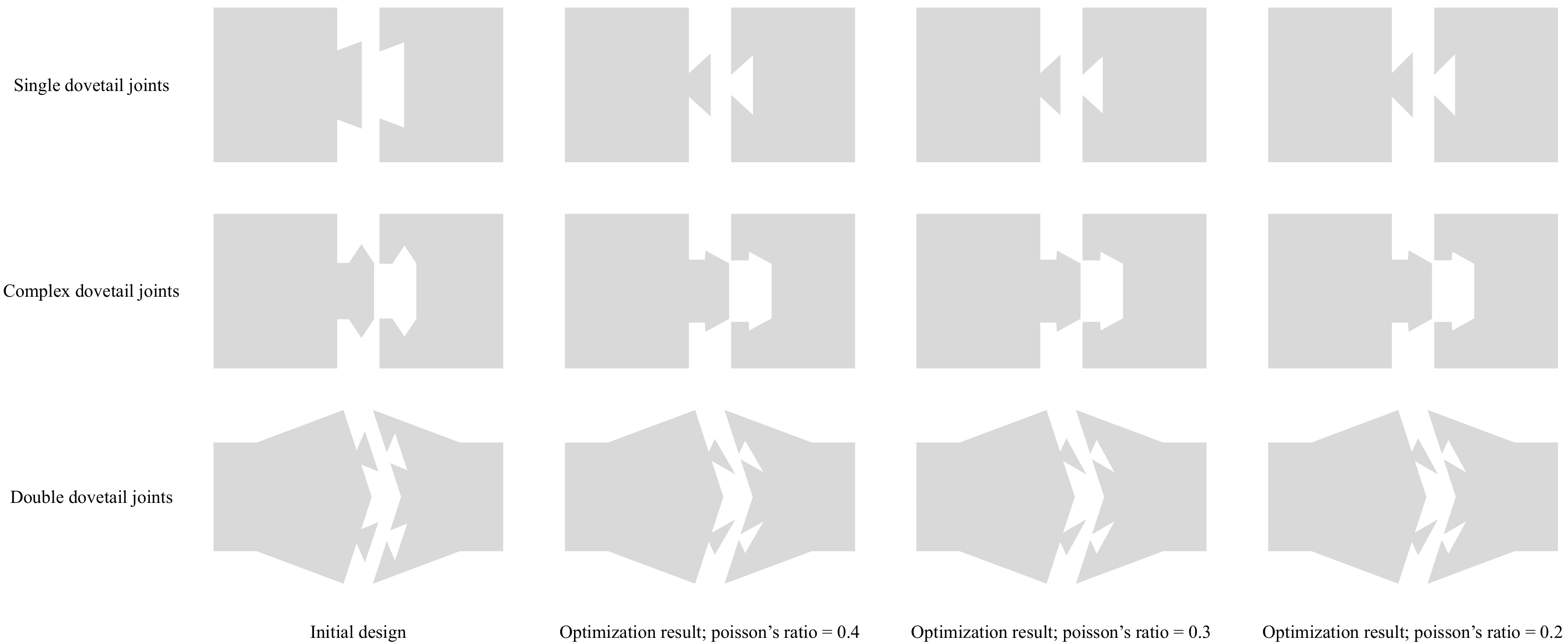}
    \caption{Optimization results from the same initial design but using different Poisson's ratios. As the difference is negligible, we observe that the optimization results are not sensitive to Poisson's ratio.}
    \label{fig:p3-poisson-test}
\end{figure*}

One interesting question raised by our method is the effect of material parameters on the optimization results.
As Young's modulus is obviously not affecting our optimization process (if simultaneously increasing the traction), we study whether changing Poisson's ratio would change the results, and the results are shown in~\fig{fig:p3-poisson-test}.
As different Poisson's ratios produce almost the same optimization result, we observe that the results are not sensitive to Poisson's ratio.

\section{Discussion}

In this project, we studied the task of dovetail joint shape optimization to maximize its stiffness, and, as existing works, we formalized the task as an optimization problem, viewing the stiffness of the joint as a function of shape design parameters.
Existing works use search algorithms, gradient-free optimizers, or surrogate models for optimization, which have limited efficiency.
To use gradient-based optimizers, we first built our own contact simulator by alternatively simulating the deformation of one side of the joint while considering the other side as rigid, using the penalty approach.
We then use the adjoint method and gradient descent for optimization.
For experiments, we first verified the gradients were correct by comparing them to the gradients calculated from the finite difference method.
We then tested the optimized joint shapes on different initializations in different dovetail shape design spaces, both synthetically and in real experiments, showing that the optimized joints are much stiffer than the initial ones.
Note that our method is not restricted to dovetail joint shape optimization but works for joints with piecewise linear joint boundaries.

We would also like to discuss some limitations and future directions of our approach.
First, the simulator has limited accuracy as several assumptions and simplifications are made, \eg, plane stress, fitting boundaries using lines for the penalty term, \etc.
Real experiments showed that the simulator is still reasonably accurate, but future work can be done on more accurate simulators.
Besides, the optimization is not in real-time, and most optimization (15 gradient descent steps) in this project takes 10 to 20 minutes to finish on a laptop computer.
However, compared to the manufacturing time, this is acceptable, and there are amortized approaches that can significantly reduce the running time.
Finally, the optimized results from gradient-based optimizers can be local minima.
Possible solutions include using different initializations, introducing randomness during optimization, \etc.

\begin{acks}

We would like to thank Eder Medina for setting up the resin 3D printer, members of the Princeton Laboratory for Intelligent Probabilistic Systems for valuable discussions and feedback, Joseph Vocaturo, and the Princeton Department of Civil and Environmental Engineering for providing the universal testing machine.
This work is partially supported by the Princeton School of Engineering and Applied Science, as well as the U.\,S.~National Science Foundation under grants \#IIS-1815070, \#IIS-2007278, and \#OAC-2118201.

\end{acks}

\bibliographystyle{ACM-Reference-Format}
\bibliography{connector_scf}

\end{document}